\documentclass[12pt,cites,graphicx]{article}
\usepackage{times}

\usepackage{epsfig}
\usepackage{overcite}

\title{Parallel Tempering: Theory, Applications, and New Perspectives}

\author{David J.\ Earl$^{a,b}$ and Michael W.\ Deem$^a$\\[3mm]
$^a$ Departments of Bioengineering and Physics \& Astronomy, 
Rice University,\\
6100 Main Street MS142, Houston, Texas 77005 USA.\\
E-mail: mwdeem@rice.edu\\[1mm]
$^b$ Rudolf Peierls Centre for Theoretical Physics, 
Oxford University,\\
1 Keble Road, Oxford OX1 3NP United Kingdom.\\
E-mail: earl@thphys.ox.ac.uk}

\begin{document}
\maketitle
\renewcommand{\thefootnote}{\fnsymbol{footnote}}

To appear in \emph{Physical Chemistry Chemical Physics}
\bigskip
\bigskip

\noindent We review the history of the parallel tempering
simulation method.  From its origins in data analysis,
the parallel tempering method has become a standard
workhorse of physiochemical simulations.
We discuss the theory behind the method and its various
generalizations.  We mention a selected set of the
many applications that have become possible with
the introduction of parallel tempering and
we suggest several promising avenues for future research.

\section{Introduction}
\label{intro}

The origins of the parallel tempering, or replica exchange, 
simulation technique can
be traced to a 1986 paper by Swendsen and Wang.\cite{SwenWang}
In this paper, a method of replica Monte Carlo was introduced in which
replicas of a system of interest are 
simulated at a series of temperatures.
Replicas at adjacent temperatures undergo a partial exchange of 
configuration information.
The more familiar form of parallel tempering with complete
exchange of configuration information
was formulated by Geyer in 1991.\cite{geyer91} 
Initially, applications of the new method were
limited to problems in statistical physics. However, following Hansmann's
use of the method in Monte Carlo simulations of a biomolecule,\cite{pt120}
Falcioni and Deem's use of parallel tempering for X-ray structure
determination,\cite{Falcioni} and Okamoto and co-worker's formulation of 
a molecular dynamics version of parallel tempering,\cite{pt133}
the use of parallel tempering in fields spanning 
physics, chemistry, biology, engineering and materials science rapidly
increased.
\bigskip


The general idea of parallel tempering is to simulate $M$
replicas of the original system of interest, each replica
typically in the canonical ensemble, and usually each replica
at a different temperature. The high temperature systems are
generally able to sample large volumes of phase space, whereas
low temperature systems, whilst having precise sampling
in a local region of phase space, may become trapped in
local energy minima during the timescale of a typical computer 
simulation. Parallel tempering achieves good sampling by 
allowing the systems at different temperatures to exchange complete
configurations. Thus, the inclusion of higher temperature
systems ensures that the lower temperature systems can
access a representative set of low-temperature regions of phase space.
This concept is illustrated in Figure 1.
\bigskip


Simulation of $M$ replicas, rather than one, requires
on the order of $M$ times more computational effort. This `extra
expense' of parallel tempering is one of the reasons for the
initially slow adoption of the method.
Eventually, it became clear that a parallel tempering
simulation is more than $1/M$ times more efficient than
a standard, single-temperature Monte Carlo simulation.
This increased efficiency derives from
allowing the lower temperature systems to sample regions of phase
space that they would not have been able to access had regular
sampling been conducted for a single-temperature simulation that was $M$ 
times as long. While not essential to the method, it is also the case
that
parallel tempering can make efficient use of large CPU clusters, where
different replicas can be run in parallel.
An additional benefit of the parallel tempering method is
the generation of
results for a range of temperatures, which may also be of interest
to the investigator. It is now widely appreciated 
that parallel tempering is a useful and powerful
computational method.
\bigskip


One of the debated issues in parallel tempering regards the details
of the exchange,
or swapping, of configurations between replicas. Pertinent
questions include how many different replicas and at
 what temperatures to use, and how frequently swaps should be attempted, and
the relative computational effort to expend on the different replicas.
Another emerging issue is how to swap only part of the system,
so as to overcome the growth as $\sqrt N$ of the number 
replicas required to simulate a system of size $N$.
  We address these points
of controversy in this review.
\bigskip


The widespread use of parallel tempering in the simulation field has 
led to the emergence of a number of new issues. 
It has also become clear that temperature may not always be the best
parameter to temper, and
parallel tempering can be conducted with order parameters other than
temperature, such as pair potentials or chemical potentials. Of interest
is how to choose the order parameter whose swapping will give the
most efficient equilibration.
It has become clear that multi-dimensional parallel tempering is
possible.
That is, swapping between a number of parameters
in the same simulation, in a multi-dimensional space of order
parameters, is feasible and sometimes advised.
The improvement in
sampling resulting from the use of parallel tempering has revealed deficiencies
in some of the most popular force fields used for atomistic simulations,
and it would seem that
the use of parallel tempering will be essential in tests of new and
improved force fields.
\bigskip

Parallel tempering can be combined with most other
simulation methods, as the exchanges, if done correctly, maintain the
detailed balance or balance condition of the underlying simulation.
Thus, there is almost an unlimited scope for the utilization of the
method in computer simulation. This leads to intriguing possibilities,
such as combining parallel tempering
with quantum methods.
\bigskip

\section{Theory}
\label{theory}

\noindent{\bf 2.1 Theory of Monte Carlo Parallel Tempering.}\\

In a typical parallel tempering simulation we have $M$ replicas,
each in the canonical ensemble, and each at a different temperature,
$T_i$. In general $T_1 < T_2 < ... < T_M$, and $T_1$ is normally the
temperature of the system of interest. Since the replicas
do not interact energetically, the partition function
of this larger ensemble is given by
\begin{equation}
Q = \prod_{i=1}^{M} \frac{q_i}{N!} \int 
                   d {\rm \bf r}_{i}^{N} 
                   \exp[-\beta_i U({\rm \bf r}_{i}^{N})]\ ,
\end{equation}
where $q_i = \prod_{j=1}^N (2 \pi m_j k_{\rm B} T_i)^{3/2}$
comes from integrating out the momenta, $m_j$ is the mass of atom
$j$, ${\rm \bf r}_{i}^{N}$
specifies the positions of the $N$ particles in system $i$, 
$\beta_i = 1 / (k_{\rm B} T_i)$ is the reciprocal temperature, and 
$U$ is the
potential energy, or the part of the Hamiltonian that does not
involve the momenta. If the probability of performing a swap move is 
equal for all conditions, exchanges between ensembles $i$ and $j$ 
are accepted with the probability
\begin{equation}
\label{eqn:accept}
A = \min\{1,~\exp\left[+(\beta_{i}-\beta_{j})(U({\rm \bf r}_{i}^{N})
           -U({\rm \bf r}_{j}^{N}))\right]\} \ .
\end{equation} 
Swaps are normally attempted between systems with adjacent temperatures,
$j=i+1$.
\bigskip

Parallel tempering is an exact method
in statistical mechanics, in that it satisfies the detailed balance
or balance condition,\cite{Deem_balance} depending on the implementation.
This is an important advantage of parallel tempering over simulated
annealing, as ensemble averages cannot be defined in the latter method.
Parallel tempering is complementary to any set of Monte Carlo moves for
a system at a single temperature, and such single-system moves are
performed between each attempted swap.  To satisfy detailed balance,
the swap moves must be performed with a certain probability, although
performing the swaps after a fixed number of single-temperature Monte
Carlo moves satisfies the sufficient condition of balance.\cite{Deem_balance}
  A typical
sequence of swaps and single-temperature Monte Carlo moves is shown
in Figure 2.
\bigskip


Kofke conducted an analysis of the average acceptance rate, 
$\langle A \rangle$, of
exchange trials and argued that this quantity should be related to the entropy 
difference between phases.\cite{pt170,pt185,Kofke2004}
For systems assumed to have Gaussian energy
distributions, typical of many systems that 
are studied using computer simulation, see Figure 3, he found the average
acceptance ratio, $\langle A \rangle$, to be given by
\begin{equation}
\langle A \rangle = {\rm erfc} \left[ \left(\frac{1}{2}C_v\right)^{1/2}\frac{1-\beta_j/\beta_i}{(1+(\beta_j/\beta_i)^{2})^{1/2}}\right] \ ,
\end{equation}
where $C_v$ is the heat capacity at constant volume, which is assumed to be 
constant in the temperature range between $\beta_i$ and $\beta_j$.
Simply put, the acceptance rate for the trials depends on the likelihood
that the system sampling the higher temperature happens to be in a
region of phase space
that is important at the lower temperature.
This theoretical analysis of the acceptance rates becomes 
useful when considering the 
optimal choice of temperatures for a parallel tempering
simulation (see Section 2.3).
\bigskip

\noindent{\bf 2.2 Theory of Molecular Dynamics Parallel Tempering.}\\
In Monte Carlo implementations of parallel tempering, we need only
consider the positions of the particles in the simulation. In
molecular dynamics, we must also take into account the momenta of all
the particles in the system. Sugita and Okamoto proposed 
a parallel tempering molecular dynamics method in which after an
exchange, the new momenta for replica $i$, $p^{(i)'}$, should be determined
as
\begin{equation}
p^{(i)'} = \sqrt{\frac{T_{\rm new}}{T_{\rm old}}}p^{(i)} \ ,
\end{equation}
where $p^{(i)}$ are the old momenta for replica $i$, and $T_{\rm old}$ and
$T_{\rm new}$ are the temperatures of the replica before and after the 
swap, respectively.\cite{pt133} This procedure
ensures the average kinetic energy remains
equal to $\frac{3}{2}N k_{\rm B} T$. The acceptance criterion for an
exchange remains the same as for the MC implementation 
(Eqn.~\ref{eqn:accept}) and satisfies detailed balance.
\bigskip

When doing parallel tempering molecular dynamics, one must take
care in the interpretation of the results. A parallel tempering 
exchange is an `unphysical' move, and so one cannot draw conclusions
about dynamics. That is, when using 
parallel tempering molecular dynamics, one is only really doing 
a form of sampling and not `true' molecular dynamics.
\bigskip

\noindent{\bf 2.3 Optimal Choice of Temperatures.}\\
How one chooses both the number of replicas employed in a parallel
tempering simulation and the temperatures of the replicas 
are questions of great importance.  One wishes to achieve the best possible
sampling with the minimum amount of computational effort.
The highest temperature must be sufficiently high so as to ensure that
no replicas become trapped in local energy minima, while the number
of replicas used must be large enough to ensure swapping occurs
between all adjacent replicas.
Several suggestions for the number and temperature of the replicas
have been offered.
It is clear from Figure 3 and Eqn.\ \ref{eqn:accept} that
the energy histograms must overlap for swaps to be accepted.
Sugita {\em et al.}\ and Kofke have proposed that the acceptance 
probability could be made
uniform across all of the different replicas, in an attempt
to ensure that each
replica spends the same amount of simulation time at each
 temperature.\cite{pt133,pt170,pt185} Kofke showed that a geometric
 progression of temperatures 
($\frac{T_i}{T_j} = {\rm constant}$) for
systems in which $C_v$ is constant across the temperatures results in 
equal acceptance ratios.
Sanbonmatsu {\em et al.}\ suggested that a target acceptance
ratio, $A_{\rm target}$, can be obtained iteratively by solving
\begin{equation}
A_{\rm target} = \exp[\Delta \beta \Delta E] \ ,
\end{equation}
where $\Delta E$ is the difference in the average energies of
the systems in adjacent temperatures.\cite{pt208} 
Iterative methods for adjusting
the temperatures of the different systems to ensure that acceptance ratios
stay within acceptable bounds had previously been proposed
and utilized by Falcioni\cite{marco_unpub} and Schug {\em et al.}\cite{pt77}
in adaptive temperature control schemes.
Rathore {\em et al.}\cite{pt90} extended these approaches to 
suggest a scheme for the optimal allocation of temperatures to
replicas that is also based on iteratively altering
system temperatures. In their scheme, the lowest temperature is
fixed, and the other system temperatures are determined
by iteratively solving
\begin{equation}
\frac{\Delta E}{\sigma_m} |_{T_j} = \left[\frac{\Delta E}{\sigma}\right]_{\rm target} \ ,
\end{equation}
for each of the temperatures, $T_j$, where 
$\sigma_m = [\sigma(T_j) + \sigma(T_i)] / 2$ is the average deviation
of the energies in the two systems. One can choose the target value
to achieve a desired acceptance ratio. 
\bigskip

Rathore {\em et al.}\ also
consider the optimal acceptance ratio and number of replicas 
in parallel tempering simulations.\cite{pt90} 
For the case studies used in their
work, they found that an acceptance ratio of 20\% yielded the best 
possible performance. That is, adding more replicas once the high
and low temperatures are fixed and the
acceptance ratio of 20\% had been achieved resulted in no increase
in the performance of the simulation.
\bigskip

Recently, Kone and Kofke have provided an analysis of the 
selection of temperature intervals in systems where $C_v$ is
assumed to be piecewise
constant across each temperature interval.\cite{Kofke2005}
They argue that although this may not always be the case, the assumption 
is reasonable and does not require an iterative scheme that
can consume valuable CPU time and which violates detailed balance.
Their analysis is based on 
maximising the mean square displacement, $\sigma^2$, of a system
as it performs the 
random walk over temperatures. 
The value of $\sigma^2$ is
proportional to the number of accepted swaps and 
$(\ln (\beta_j / \beta_i))^{2}$. By maximizing $\sigma^2$  with respect 
to the acceptance probability, they found that an acceptance probability
of 23\% is optimal. This value  is strikingly similar to the
empirically determined 20\% of Rathore {\em et al.}
Kone and Kofke suggest ``tuning'' the temperature intervals
to achieve the 23\% acceptance probability
during the initial equilibration of a simulation.
This approach appears to be an efficient method to select
temperature intervals in parallel tempering simulations that
mix efficiently.
\bigskip

A similar scheme for choosing the temperatures has
recently been proposed by Katzgraber {\em et al.}, which uses an
adaptive feedback-optimized algorithm to minimize round-trip
times between the lowest and highest temperatures.\cite{ccp2004}
This approach more directly characterizes the mixing between
the high and low temperature systems.  In complex cases, where
there are subtle bottlenecks in the probability of exchange
of configurations, the round-trip time is likely to better characterize
the overall efficiency of parallel tempering than is the average
acceptance probability.  The approach of Katzgraber {\em et al.}\
is a promising one for such complex cases.
\bigskip

A related issue is how much simulation effort should be
expended on each replica.  
For example, it would seem that the low temperature replicas
would benefit from additional simulation effort, as the
correlation times at lower temperature are longer.  This issue
is untouched in the literature.
\bigskip

Since the width of the energy histograms increases as $\sqrt N$,
but the average energy increases as $N$, 
the number of replicas increases as $\sqrt N$,
where $N$ is is the system size.\cite{Falcioni}  One, therefore, would
like a method where only part of the configurational
degrees of freedom are exchanged.  Interestingly, this issue
was solved in Swendsen and Wang's 1986 paper for spin systems,\cite{SwenWang} 
but it has not been solved in an exact, efficient way for atomistic systems.
The main difficulty seems to be in defining a piece of a system
that can be exchanged without the penalty of a large surface energy.
\bigskip

\noindent{\bf 2.4 Parallel Tempering with Alternative Parameters and 
Sampling Methods.}\\
The general idea of parallel tempering is not limited to exchanges
or swaps between systems at different temperatures. Investigators
have developed a number of methods based on swapping alternative
parameters in order to minimize barriers that inhibit correct sampling.
Additionally, parallel tempering can be combined with a large number
of alternative sampling methods, and its use has led to a great
improvement in the sampling of many existing computational methods. 
\bigskip

Fukunishi {\em et al.}\ developed a Hamiltonian parallel tempering
method that they applied to biomolecular systems.\cite{pt167} In this 
approach, only part of the interaction energy between particles is scaled
between the different
replicas. In their work, they conduct case studies using two
different implementations of their approach. In the first they scale 
hydrophobic interactions between replicas. In the second, they
scale the van der Waals interactions between replicas by introducing
a cut-off in the interaction, effectively allowing chains of atoms
to pass through each other. The acceptance probability in Hamiltonian
parallel tempering for a swap between replicas $i$ and $j$ is given by
\begin{equation}
\label{eqn:Ham}
A = \min\{1,~\exp\left[-\beta \left( \left[ H_{i}(X') + H_{j}(X) \right] - \left[ H_{i}(X) + H_{j}(X') \right] \right) \right]\} \ ,
\end{equation}
where $H_{i}(X)$ is the Hamiltonian of configuration $X$ in replica $i$,
and configurations $X$ and $X'$ are the configurations in replicas $i$
and $j$, respectively, prior to the swap attempt. 
\bigskip


Parallel tempering using multiple swapping variables was first proposed
and developed by Yan and de Pablo.\cite{pt151,pt154} Instead of considering
a one-dimensional array of replicas at different temperatures, 
they suggested using an $n$-dimensional array, where each dimension
represented a parameter that varied between replicas. Swaps both
within and between dimensions were allowed in their scheme. In their
first work they conducted parallel tempering between different temperatures
and chemical potentials in the grand canonical ensemble, but the
scheme they proposed was general. They showed that extensions of 
parallel tempering to multiple dimensions are limited only by the
imagination of the investigator in choosing the variables to swap 
and the available computational resources.
Sugita {\em et al.}\ utilized multdimensional exchanges in molecular
dynamics studies.\cite{pt156}
\bigskip

de Pablo and co-workers also implemented parallel tempering
in the multicanonical
ensemble.\cite{pt165} In the multicanonical ensemble, the
probability distribution is no longer Boltzmann, but becomes
\begin{equation}
p({\bf r}^N) = ({\rm const}) e^{-\beta U({\bf r}^N)} w({\bf r}^N)
\label{e8}
\end{equation}
The weight factors, $w({\bf r}^N)$, are chosen so as
to lower the barriers in the system.
de Pablo and co-workers derived multicanonical weights by an iterative
process using a Boltzmann inversion of histograms. Another way
to write Eq.\ (\ref{e8}) is to use instead of
the Hamiltonian $U$, the weighted Hamiltonian $U+\xi(U)$
when attempting swap moves, where $\xi(U)$ is an umbrella 
potential. By using a multicanonical ensemble, de Pablo and co-workers
were able to
reduce the number of replicas required in their simulation, due to
a broader overlap of thermodynamic-property histograms.
In general, when combined with a multicanonical simulation, a short 
parallel tempering
run can be performed, and the multicanonical weight factors can be determined
by using histogram reweighting. These weights can then be used in the
multicanonical part of the calculation.\cite{multi1} Parallel tempering can
be combined with a multicanonical simulation.  That is, in the 
multicanonical simulation, a number of replicas, each in the multicanonical
ensemble but each with different multicanonical weight factors covering 
different energy ranges, may be employed.\cite{multi1} 
It should be noted that far
fewer replicas are needed in this method than in typical parallel
tempering because the energy ranges
covered in a multicanonical simulation are far wider than in a 
canonical simulation. The weight factors utilized in these methods may
then be iteratively improved during the equilibration period
as the simulation proceeds, using histogram
reweighting techniques. 
\bigskip

In free energy perturbation calculations, a parameter
$\lambda$ is introduced. One wishes to compute the free energy difference
to go from an initial ($\lambda = 0$) state and a
final ($\lambda = 1$) state. 
For parallel tempering with free energy perturbation one
can consider $M$ replicas, each with a different $\lambda$ parameter, 
where each replica has a slightly different Hamiltonian
\begin{equation}
U_{\lambda} = U_{\lambda =0} + \lambda (U_{\lambda =1} - U_{\lambda =0}) \ .
\end{equation}
Swaps may be attempted between replicas using the Hamiltonian acceptance
criterion (Eq.~\ref{eqn:Ham}), and the free energy difference between
two lambda parameters can be determined  as in regular free energy 
calculations. Of course, one may utilize a number of different temperature
replicas for each value of $\lambda$ in a multidimensional approach.
Use of parallel tempering
in multicanonical simulations,
 free energy calculations, and umbrella sampling 
is growing.\cite{pt156,pt165,multi1,multi2,multi3,pt125,pt201}
\bigskip

One of the most fruitful combinations of parallel tempering with
existing sampling techniques has been with density of states methods
based on Wang-Landau sampling.\cite{WangLandau} Density of states methods
are similar to multicanonical ones in that the weight factor is the
reciprocal of the density of states. However, in density of states methods
a random walk in energy space is conducted, and a running estimate of
the inverse of the density of states as a function of the energy
is performed. Alternatively the configurational temperature is
collected as a function of the energy and the density of states 
determined by integrating the inverse temperature.\cite{dePablo6} Other
sets of conjugate variables can also profitably be used.\cite{Swen2004} 
These methods effectively circumvent the tedious and
time consuming process of calculating weight factors in multicanonical
simulations. de Pablo and co-workers have proposed extended ensemble
density of states methods where overlapping windows or replicas of 
different energy or reaction/transition coordinate values are 
utilized.\cite{dePablo1} Configurational swaps between windows are attempted
at regular intervals to prevent the simulations in the parallel 
replicas from becoming stuck in non-representative regions of phase
space. A combination of density of states methods and parallel
tempering has successfully been used to study protein 
folding\cite{dePablo2,dePablo3,dePablo4} and solid-liquid 
equilibria.\cite{dePablo5}
\bigskip


Vlugt and Smit applied parallel tempering to the transition path sampling
method.\cite{pt94}
They showed that parallel tempering conducted between different temperatures
and between different regions along transition paths is
able to overcome the problem of multiple saddle points on a free
energy surface.  Parallel tempering transition path sampling
can provide for more accurate estimates of
transition rates between stable states than
single-temperature Monte Carlo transition path sampling.
\bigskip


Parallel tempering has been combined with a number of other 
computational methods, and in almost all cases its use has
resulted in better sampling and an increase in the accuracy
of the computational method. Prominent examples include parallel
tempering with cavity bias to study the phase diagram of Lennard-Jones
fluids,\cite{pt130} with analytical rebridging for the simulation
of cyclic peptides,\cite{pt204} and with the wormhole algorithm
to explore the phase behavior of random copolymer melts.\cite{pt24}
\bigskip

Very recently an extension to parallel tempering, known as 
{\em Virtual-Move Parallel Tempering}, has been proposed by
Coluzza and Frenkel.\cite{VMPT} In their scheme they include 
information about all possible parallel tempering moves between all 
replicas in the system, rather than just between adjacent replicas,
when accumulating statistical averages. This approach is essentially
a parallel tempering version of the ``waste recyling" Monte Carlo 
method of Frenkel \cite{Frenkel_waste} and has been shown to improve
statistical averaging by upto a factor of 20.
\bigskip

\noindent{\bf 2.5 Non-Boltzmann Distributions.}\\
Since their introduction in the late 1980s, Tsallis statistics
have become increasingly important in 
statistical mechanics.\cite{Tsallis} Due to their power-law, 
rather than Boltzmann, properties,
Tsallis statistics generally lead to smaller energy barriers.
Therefore, optimization with Tsallis, rather than  Boltzmann, statistics
can be very useful in energy minimization problems.
Whitfield
{\em et al.}\ have developed a version of the parallel tempering
algorithm
that is based upon Tsallis statistics.\cite{pt37}
This method has been used, for example, for  
fast conformational searches of peptide molecules.\cite{pt50}
\bigskip

\section{Applications}
\label{apps}

\noindent{\bf 3.1 Polymers.}\\
Simulations of polymeric systems are notoriously difficult due
to chain tangling, the high density of the systems of interest, 
and the large system sizes required to accurately model high
molecular weight species.
The first application of parallel tempering to polymeric
systems was by Yan and de Pablo to high molecular weight species\cite{pt154}.
Bunker and Dunweg\cite{pt71} were the first to utilize excluded
volume parallel tempering, where different replicas have different
core potentials. They studied polymer melts for polymer chain lengths
ranging from 60 to 200 monomers. Their method created a thermodynamic
path from the full excluded volume system to an ideal gas of random
walks and increased the efficiency of all their simulations. 
Bedrov and Smith\cite{pt158} studied fully atomistic polymer
melts of 1,4-polybutadiene at a range of temperatures, performing
parallel tempering
swaps isobarically. They showed that their parallel tempering
approach provided
a substantial improvement in equilibration and sampling of conformational 
phase space when compared to regular MD simulations. See Figure 4.
Theodorou and co-workers studied {\em cis}-1,4 
polyisoprene melts using parallel tempering
and once again found that use of parallel tempering
resulted in far quicker equilibration over a range of temperatures.\cite{pt159}
More recently, Banaszak {\em et al.} have utilized hyperparallel tempering
in an osmotic ensemble to study the solubility of ethylene in low-density
polyethylene.\cite{Banaszak} Using their novel method they were able to
examine the effect of both polyethylene chain length and branching on
the solubility of ethylene.
\bigskip

\noindent{\bf 3.2 Proteins.}\\
Biological systems, particularly proteins, are 
computationally challenging because they have particularly rugged
energy landscapes that are difficult for regular Monte Carlo
and molecular dynamics techniques to traverse.
Hansmann was the first to apply parallel tempering
to biological molecules
in a Monte Carlo based study of the simple  7-amino acid
Met-enkephalin peptide.\cite{pt120}
 Hansmann showed that parallel tempering based simulations could 
overcome the ``simulation slowdown'' problem and were more
efficient than regular canonical Monte Carlo simulations.
The application of parallel tempering to biological problems, however, did not
take-off until Sugita and Okamoto's work that introduced the
use of molecular dynamics parallel tempering.\cite{pt133}
They applied their approach to 
Met-enkephalin and demonstrated that their parallel tempering based method did 
not get trapped in local energy minima, unlike regular microcanonical
molecular dynamics simulations of the same molecule.
\bigskip

Following demonstration of the power of parallel tempering
for molecular systems,
its use in the biological simulation community rapidly 
expanded.
Parallel tempering has been used to determine folding free energy contour maps
for a number of proteins, revealing details about folding mechanisms
and intermediate state structures\cite{pt1,pt54,pt55,pt56,pt59}
and has facilitated the simulation 
of membrane proteins.\cite{pt198,pt99,pt184,pt122}.
Parallel tempering has proved to be particularly powerful when applied to 
NMR structure refinement and in the interpretation of data
from NMR,\cite{pt107,pt129,pt142,pt144,pt102}
circular dichroism,\cite{pt103}
IR spectra,\cite{pt61}
and electric deflection data\cite{pt143}
of proteins and peptides.
For models of globular proteins and oligomeric peptides,
parallel tempering has been used to study previously unexplored regions
of phase diagrams and to sample aggregate transitions.\cite{pt188,pt189}
In the study of sucrose solutions near the glass transition 
temperature, parallel tempering
 simulations showed a much better fit to experimental
data than did conventional NPT MC results.\cite{pt5}.
Other interesting work using parallel tempering
includes studies 
of the thermodynamics of fibril formation using an intermediate 
resolution protein model\cite{pt104} and of the hypervariable
regions of an antibody domain where the chief interactions 
governing conformational equilibria in these systems were
determined.\cite{pt108}.
\bigskip

With this increased sampling ability of parallel tempering 
has come the realization that
current force fields for biological simulation are lacking
in some respects.
Parallel tempering simulations of solvated biological 
molecules have also revealed deficiencies in popular implicit solvent
models.\cite{pt58,pt59,pt83}  As parallel tempering can also be used with 
explicit solvent models, the differences between the treatments
can be determined, and in the future such simulations
could be used to improve 
implicit solvent models.
\bigskip

Brooks and co-workers have developed a multiscale modeling toolkit
that can interface with the popular CHARMM and AMBER molecular
simulation codes.\cite{pt200}  Parallel tempering is implemented in the toolkit
to allow enhanced sampling and is used to study the {\em ab initio}
folding of peptides from first principles. Parallel tempering
has clearly become
the method of choice in {\em ab initio} protein folding as evidenced 
by the the work of Skolnick and co-workers,\cite{pt53} 
Garcia and Sanbonmatsu,\cite{pt208,pt209,pt210}  and Yang
{\em et al.}\cite{pt199}
\bigskip

When examining the biological and chemical literature of
parallel tempering, it is apparent that the vast majority 
of work is based on molecular dynamics, rather than Monte Carlo.
As one is not doing `true' MD when using parallel tempering,
 there is no reason 
why Monte Carlo methodologies cannot be implemented more frequently
in the biological and chemical communities.  Indeed, we expect
this to be a promising avenue for future research.
\bigskip
 
\noindent{\bf 3.3 Solid state.}\\
Crystal structure solution provided one of the first mainstream 
atomistic simulation examples of the power of parallel tempering.
Falcioni and Deem used parallel tempering in a biased MC scheme
to determine the structures of zeolites from powder diffraction
data.\cite{pt150} For complicated zeolite structures containing more than
eight unique tetrahedral atoms, simulated annealing is unable to solve 
the crystal structures. However, parallel tempering
simulations were shown to be able 
to solve the structures of all complex zeolites, including the most
complicated zeolite structure, ZSM-5, which contains twelve unique
tetrahedral atoms. ZefsaII has since been successfully used to solve 
the structures of at least a dozen newly synthesized zeolites 
and is freely downloadable on the web.
 A similar approach to crystal structure determination
from powder diffraction data has been implemented by Favre-Nicolin {\em 
et al.},\cite{pt84} and this method has been successful in solving
several structures.\cite{pt85,pt86,pt87}
\bigskip

A seminal simulation study of the rate of crystal nucleation 
by Auer and Frenkel utilized the parallel tempering method by allowing swaps
between `windows' at different points along the reaction co-ordinate 
from the liquid to solid state.\cite{Frenkel} This work
introduced, for the first time, the ability to calculate
nucleation rates from first principles.
\bigskip

Other examples of solid-state parallel tempering simulations
include the computation of 
sodium ion distributions in zeolites,\cite{pt12} studying the finite 
temperature behavior of C$_{60}$ clusters,\cite{pt6} the 
simulation of Si surfaces,\cite{pt43,pt88} and the explanation
of the titration behavior of MbCO over a range of pH values.\cite{pt96}.
\bigskip

\noindent{\bf 3.4 Spin glass.}\\
Spin glasses have provided a severe test of the effectiveness
of parallel tempering.\cite{marinari98}
In the Parisi solution of the infinite range
Edwards-Anderson model, widely believed by many but not all physicists
to apply to finite-range spin glasses as well, there is a finite
energy for excitations above the ground state,
 and the boundary of these excitations has
a space-filling structure.  Initial simulations for the Edwards-Anderson
model confirmed the finite excitation energy.\cite{pt67}
Initial suggestions for a fractal surface\cite{pt67} were
ruled out by later simulations.\cite{pt40}
For the vector spin glass model, the excitation energy
was again found to be finite.\cite{pt39}
Initial suggestions of a fractal surface were also
largely ruled out in later simulations.\cite{pt116}
\bigskip

\noindent{\bf 3.5 Quantum.}\\
Quantum level systems, whilst being far more computationally demanding
than classical systems, may benefit from the improved sampling
provided by parallel tempering.
So far, the main application of parallel tempering at the 
quantum level has been in studies of phase transitions and in
the location of energy minima in complex systems. Parallel tempering
is ideal for these
studies, as dynamics are not of interest.
Okamoto and co-workers conducted parallel tempering based
{\em ab initio} correlated electronic structure calculations.\cite{pt121}
In their studies of Li clusters, they demonstrated that parallel tempering
could be 
successfully applied to systems described with a high level of detail.
Sengupta {\em et al.}\ combined quantum Monte Carlo with parallel tempering
to study the phase 
diagram of a 1-D Hubbard model.\cite{pt38} Quantum parallel tempering
was found to
significantly reduce ``sticking'' effects, where the simulation gets
stuck in the incorrect phase close to the phase boundary.
\bigskip
 
Shin {\em et al.}\ have studied quantum phase transitions of water
clusters,\cite{pt2} where the rotational modes can be highly quantum.
Parallel tempering
allowed for efficient conformational sampling.
 They remark that ``combining Car-Parrinello approach
with replica exchange [parallel tempering] and path 
integral molecular dynamics
can provide an ideal methodology for studying quantum behavior of
clusters.'' Although the suggested approach 
is highly computationally expensive, it may become increasingly feasible
in future years.
Parallel tempering has also been successfully employed in a study of the
finite temperature optical spectroscopy of CaAr$_n$ clusters
\cite{pt140} and in quantum path integral
simulations of the
solid-liquid phase diagrams of Ne$_{13}$ and (para-H$_2$)$_{13}$
clusters.\cite{pt168}
\bigskip

\noindent{\bf 3.6 General Optimization Problems.}\\
Parallel tempering
has been successfully used in a number of general optimization
problems. Habeck {\em et al.}\ developed a sampling algorithm for
the exploration of probability densities that arise in Bayesian data
analysis.\cite{pt52} Their approach utilized Tsallis 
statistics, and the effectiveness of parallel tempering
was demonstrated by interpreting
experimental NMR data for a folded protein.
In image analysis, parallel tempering
has been shown to
lead to an improvement by a factor of two for
both success rate and mean position error when compared to simulated
annealing approaches.\cite{pt141}
Parallel tempering
has also been utilized to locate the global minima of complex and
rugged potential energy surfaces that arise in atomistic models
of receptor-ligand docking \cite{pt123} and in risk analysis.\cite{pt136}
\bigskip

\section{Conclusion}
\label{conclusions}
In this review we have given an overview of the history
of parallel tempering.  We have described the basic theory and
many of the extensions to the original method.
Several examples in a variety of physiochemical arenas
have been discussed.
Highlighted technical aspects to sort out include best allocations to cluster
computers,\cite{EarlDeem} determination of the optimal
amount of simulation effort to expend
on each replica, and partial swapping of partial configuration information
for atomistic systems.
\bigskip

A number of potential new areas for application of parallel
tempering occur to us.  One rather large one is the application of
parallel tempering, rather than simulated annealing,\cite{XPLOR} to X-ray
single-crystal structure solution.  A related issue is the 
prediction of polymorphs for crystals of small, organic drug molecules.
Also related is use of parallel tempering in rational drug design---most
current approaches use grid searching, traditional Monte Carlo, or at
best simulated annealing.\cite{Shakhnovich}
Another physical application where enhanced sampling might be of use is in
field theories for polymeric systems with non-trivial phase
structure.\cite{Fredrickson}  Also possible would be the complementary
inclusion in \emph{ab initio} molecular dynamics, if sampling only is
desired.  Even experimental applications could be possible in
materials discovery\cite{pt203} or 
laboratory protein evolution\cite{BogaradDeem}.

\bigskip

\clearpage

\bibliography{pt_review}

\begin{thebibliography}{100}

\bibitem{SwenWang}
R.~H. Swendsen and J.-S. Wang, {\em Phys. Rev. Lett.}, 1986, {\bf 57}, 2607.

\bibitem{geyer91}
C.~J. Geyer In {\em Computing Science and Statistics: Proceedings of the 23rd
  Symposium on the Interface}, p. 156, New {York}, 1991. American Statistical
  Association.

\bibitem{pt120}
U.~H.~E. Hansmann, {\em Chem. Phys. Lett.}, 1997, {\bf 281}, 140.

\bibitem{Falcioni}
M.~Falcioni and M.~W. Deem, {\em J. Chem. Phys.}, 1999, {\bf 110}, 1754.

\bibitem{pt133}
Y.~Sugita and Y.~Okamoto, {\em Chem. Phys. Lett.}, 1999, {\bf 314}, 141.

\bibitem{Deem_balance}
V.~I. Manousiouthakis and M.~W. Deem, {\em J. Chem. Phys.}, 1999, {\bf 110},
  2753.

\bibitem{pt170}
D.~A. Kofke, {\em J. Chem. Phys.}, 2002, {\bf 117}, 6911.

\bibitem{pt185}
D.~A. Kofke, {\em J. Chem. Phys.}, 2004, {\bf 120}, 10852.

\bibitem{Kofke2004}
D.~A. Kofke, {\em J. Chem. Phys.}, 2004, {\bf 121}, 1167.

\bibitem{pt208}
K.~Y. Sanbonmatsu and A.~E. Garcia, {\em Proteins}, 2002, {\bf 46}, 225.

\bibitem{marco_unpub}
M.~Falcioni, unpublished.

\bibitem{pt77}
A.~Schug, {\em Proteins}, 2004, {\bf 57}, 792.

\bibitem{pt90}
N.~Rathore, M.~Chopra, and J.~J. de~Pablo, {\em J. Chem. Phys.}, 2005, {\bf
  122}, 024111.

\bibitem{Kofke2005}
A.~Kone and D.~A. Kofke, {\em J. Chem. Phys.}, 2005, {\bf 122}, 206101.

\bibitem{ccp2004}
D.~Huse, H.~G. Katzgraber, S.~Trebst, and M.~Troyer In {\em Europhysics
  Conference on Computational Physics 2004, Book of Abstracts}, p.~63. European
  Physical Society, 2004.

\bibitem{pt167}
H.~Fukunishi, O.~Watanabe, and S.~Takada, {\em J. Chem. Phys.}, 2002, {\bf
  116}, 9058.

\bibitem{pt151}
Q.~Yan and J.~J. de~Pablo, {\em J. Chem. Phys.}, 1999, {\bf 111}, 9509.

\bibitem{pt154}
Q.~Yan and J.~J. de~Pablo, {\em J. Chem. Phys.}, 2000, {\bf 113}, 1276.

\bibitem{pt156}
Y.~Sujita, A.~Kitao, and Y.~Okamoto, {\em J. Chem. Phys.}, 2000, {\bf 113},
  6042.

\bibitem{pt165}
R.~Faller, Q.~Yan, and J.~J. de~Pablo, {\em J. Chem. Phys.}, 2002, {\bf 116},
  5419.

\bibitem{multi1}
Y.~Sugita and Y.~Okamoto, {\em Chem. Phys. Lett.}, 2000, {\bf 329}, 261.

\bibitem{multi2}
A.~Mitsutake, Y.~Sugita, and Y.~Okamoto, {\em J. Chem. Phys.}, 2003, {\bf 118},
  6664.

\bibitem{multi3}
A.~Mitsutake, Y.~Sugita, and Y.~Okamoto, {\em J. Chem. Phys.}, 2003, {\bf 118},
  6676.

\bibitem{pt125}
K.~Murata, Y.~Sugita, and Y.~Okamoto, {\em Chem. Phys. Lett.}, 2004, {\bf 385},
  1.

\bibitem{pt201}
Y.~Okamoto, {\em J. Molec. Graph. Mod.}, 2004, {\bf 22}, 425.

\bibitem{WangLandau}
F.~Wang and D.~Landau, {\em Phys. Rev. Lett.}, 2001, {\bf 86}, 2050.

\bibitem{dePablo6}
Q.~Yan and J.~J. de~Pablo, {\em Phys. Rev. Lett.}, 2003, {\bf 90}, 035701.

\bibitem{Swen2004}
M.~Fasnacht, R.~H. Swendsen, and J.~M. Rosenberg, {\em Phys. Rev. E}, 2004,
  {\bf 69}, 056704.

\bibitem{dePablo1}
N.~Rathore and J.~J. de~Pablo, {\em J. Chem. Phys.}, 2002, {\bf 117}, 7781.

\bibitem{dePablo2}
N.~Rathore, T.~A. Knotts, and J.~J. de~Pablo, {\em J. Chem. Phys.}, 2003, {\bf
  118}, 4285.

\bibitem{dePablo3}
N.~Rathore, T.~A. Knotts, and J.~J. de~Pablo, {\em Biophysical J.}, 2003, {\bf
  85}, 3963.

\bibitem{dePablo4}
N.~Rathore, Q.~Yan, and J.~J. de~Pablo, {\em J. Chem. Phys.}, 2004, {\bf 120},
  5781.

\bibitem{dePablo5}
E.~A. Mastny and J.~J. de~Pablo, {\em J. Chem. Phys.}, 2005, {\bf 122}, 124109.

\bibitem{pt94}
T.~J.~H. Vlugt and B.~Smit, {\em Chem. Phys. Comm.}, 2001, {\bf 2}, 1.

\bibitem{pt130}
V.~Ortiz, J.~R. Maury-Evertsz, and G.~E. Lopez, {\em Chem. Phys. Lett.}, 2003,
  {\bf 368}, 452.

\bibitem{pt204}
M.~G. Wu and M.~W. Deem, {\em J. Chem. Phys.}, 1999, {\bf 111}, 6625.

\bibitem{pt24}
J.~Houdayer and M.~Muller, {\em Macromolecules}, 2004, {\bf 37}, 4283.

\bibitem{VMPT}
I.~Coluzza and D.~Frenkel, submitted.

\bibitem{Frenkel_waste}
D.~Frenkel, {\em Proc. Natl. Acad. Sci. USA}, 2004, {\bf 101}, 17571.

\bibitem{Tsallis}
C.~Tsallis, {\em J. Stat. Phys.}, 1988, {\bf 52}, 479.

\bibitem{pt37}
T.~M. Whitfield, L.~Bu, and J.~E. Straub, {\em Physica A}, 2002, {\bf 305},
  157.

\bibitem{pt50}
S.~Jang, S.~Shin, and Y.~Pak, {\em Phys. Rev. Lett.}, 2003, {\bf 91}, 058305.

\bibitem{pt71}
A.~Bunker and B.~Dunweg, {\em Phys. Rev. E}, 2000, {\bf 63}, 016701.

\bibitem{pt158}
D.~Bedrov and G.~D. Smith, {\em J. Chem. Phys.}, 2001, {\bf 115}, 1121.

\bibitem{pt159}
M.~Dozastakis, V.~G. Mavrantzas, and D.~N. Theodorou, {\em J. Chem. Phys.},
  2001, {\bf 115}, 11352.

\bibitem{Banaszak}
B.~J. Banaszak, R.~Faller, and J.~J. de~Pablo, {\em J. Chem. Phys.}, 2004, {\bf
  120}, 11304.

\bibitem{pt1}
R.~Zhou, {\em J. Molec. Graph. Mod.}, 2004, {\bf 22}, 451.

\bibitem{pt54}
R.~Zhou and B.~J. Berne, {\em Proc. Natl. Acad. Sci. USA}, 2002, {\bf 99},
  12777.

\bibitem{pt55}
R.~Zhou, {\em Proc. Natl. Acad. Sci. USA}, 2003, {\bf 100}, 13280.

\bibitem{pt56}
A.~E. Garcia and J.~E. Onuchic, {\em Proc. Natl. Acad. Sci. USA}, 2003, {\bf
  100}, 13898.

\bibitem{pt59}
R.~Zhou, B.~J. Berne, and R.~Germain, {\em Proc. Natl. Acad. Sci. USA}, 2001,
  {\bf 98}, 14931.

\bibitem{pt198}
W.~Im and C.~L.~B. III, {\em J. Mol. Bio.}, 2004, {\bf 337}, 513.

\bibitem{pt99}
W.~Im, M.~Feig, and C.~L.~B. III, {\em Biophysical J.}, 2003, {\bf 85}, 2900.

\bibitem{pt184}
H.~Kokubo and Y.~Okamoto, {\em J. Chem. Phys.}, 2004, {\bf 120}, 10837.

\bibitem{pt122}
H.~Kokubo and Y.~Okamoto, {\em Chem. Phys. Lett.}, 2004, {\bf 392}, 168.

\bibitem{pt107}
T.~Haliloglu, A.~Kolinski, and J.~Skolnick, {\em Biopolymers}, 2003, {\bf 70},
  548.

\bibitem{pt129}
G.~L. Penna, A.~Mitsutake, M.~Masayu, and Y.~Okamoto, {\em Chem. Phys. Lett.},
  2003, {\bf 380}, 609.

\bibitem{pt142}
J.~Chen, H.-S. Won, W.~Im, H.~J. Dyson, and C.~L.~B. III, {\em J. Biomolec.
  NMR}, 2005, {\bf 31}, 59.

\bibitem{pt144}
J.~Chen, W.~Im, and C.~L.~B. III, {\em J. Am. Chem. Soc.}, 2004, {\bf 126},
  16038.

\bibitem{pt102}
J.~Vreede, W.~Crielaard, K.~J. Hellingwerf, and P.~G. Bolhuis, {\em Biophysical
  J.}, 2005, {\bf 88}, 3525.

\bibitem{pt103}
G.~S. Jas and K.~Kuczera, {\em Biophysical J.}, 2004, {\bf 87}, 3786.

\bibitem{pt61}
S.~Gnanakaran, R.~M. Hochstrasser, and A.~E. Garcia, {\em Proc. Natl. Acad.
  Sci. USA}, 2004, {\bf 101}, 9229.

\bibitem{pt143}
P.~Dugourd, R.~Antoine, G.~Breaux, M.~Broyer, and M.~F. Jarrold, {\em J. Am.
  Chem. Soc.}, 2005, {\bf 127}, 4675.

\bibitem{pt188}
D.~L. Pagan, M.~E. Grachava, and J.~D. Gunton, {\em J. Chem. Phys.}, 2004, {\bf
  120}, 8292.

\bibitem{pt189}
M.~Cecchini, F.~Rao, M.~Seeber, and A.~Caflisch, {\em J. Chem. Phys.}, 2004,
  {\bf 121}, 10748.

\bibitem{pt5}
N.~C. Ekdawi-Sever, P.~B. Conrad, and J.~J. de~Pablo, {\em J. Phys. Chem. A},
  2001, {\bf 105}, 734.

\bibitem{pt104}
H.~D. Nguyen and C.~K. Hall, {\em Biophysical J.}, 2004, {\bf 87}, 4122.

\bibitem{pt108}
M.~K. Fenwick and F.~A. Escobedo, {\em Biopolymers}, 2002, {\bf 68}, 160.

\bibitem{pt58}
H.~Nymeyer and A.~E. Garcia, {\em Proc. Natl. Acad. Sci. USA}, 2003, {\bf 100},
  13934.

\bibitem{pt83}
R.~Zhou, {\em Proteins}, 2003, {\bf 53}, 148.

\bibitem{pt200}
M.~Feig, J.~Karanicolas, and C.~L.~B. III, {\em J. Molec. Graph. Mod.}, 2004,
  {\bf 22}, 377.

\bibitem{pt53}
D.~Kihara, H.~Lu, A.~Kolinski, and J.~Skolnick, {\em Proc. Natl. Acad. Sci.
  USA}, 2001, {\bf 98}, 10125.

\bibitem{pt209}
A.~E. Garcia and K.~Y. Sanbonmatsu, {\em Proteins}, 2001, {\bf 42}, 345.

\bibitem{pt210}
K.~Y. Sanbonmatsu and A.~E. Garcia, {\em Proc. Natl. Acad. Sci. USA}, 2002,
  {\bf 99}, 2782.

\bibitem{pt199}
W.~Y. Yang, J.~W. Pitera, W.~C. Swope, and M.~Gruebele, {\em J. Mol. Bio.},
  2004, {\bf 336}, 241.

\bibitem{pt150}
M.~Falcioni and M.~W. Deem, {\em J. Chem. Phys.}, 1999, {\bf 110}, 1754.

\bibitem{pt84}
V.~Favre-Nicolin and R.~Cerny, {\em J. Appl. Cryst.}, 2002, {\bf 35}, 734.

\bibitem{pt85}
M.~C. Garcia-Cuesta, A.~M. Lozano, J.~J. Melendez-Martinez, F.~Luna-Giles,
  A.~L. Ortiz, L.~M. Gonzalez-Mendez, and F.~L. Cumbera, {\em J. Appl. Cryst.},
  2004, {\bf 37}, 993.

\bibitem{pt86}
P.~Y. Zavalij, S.~Yang, and M.~S. Whittingham, {\em Acta Cryst.}, 2003, {\bf
  B59}, 753.

\bibitem{pt87}
E.~Dova, R.~Peschar, M.~Sakata, K.~Kato, A.~F. Stassen, H.~Schenk, and J.~G.
  Haasnoot, {\em Acta Cryst.}, 2004, {\bf B60}, 528.

\bibitem{Frenkel}
S.~Auer and D.~Frenkel, {\em Nature}, 2001, {\bf 409}, 1020.

\bibitem{pt12}
C.~Beauvais, X.~Guerrault, F.-X. Coudert, A.~Boutin, and A.~H. Fuchs, {\em J.
  Phys. Chem. B}, 2004, {\bf 108}, 399.

\bibitem{pt6}
F.~Calvo, {\em J. Phys. Chem. B}, 2001, {\bf 105}, 2183.

\bibitem{pt43}
C.~V. Ciobanu and C.~Predescu, {\em Phys. Rev. B}, 2004, {\bf 70}, 085321.

\bibitem{pt88}
F.~C. Chuang, C.~V. Ciobanu, C.~Predescu, C.~Z. Wang, and K.~M. Ho, {\em Surf.
  Sci.}, 2005, {\bf 578}, 183.

\bibitem{pt96}
B.~Rabenstein and E.-W. Knapp, {\em Biophysical J.}, 2001, {\bf 80}, 1141.

\bibitem{marinari98}
E.~Marinari, G.~Parisi, and J.~Ruiz-{Lorenzo} in {\em Spin Glasses and Random
  Fields}, ed. A.~Young, Vol. ~12 of {\em Directions in Condensed Matter
  Physics;}
\newblock World Scientific, Singapore, 1998.

\bibitem{pt67}
H.~G. Katzgraber, M.~Palassini, and A.~P. Young, {\em Phys. Rev. B}, 2001, {\bf
  63}, 184422.

\bibitem{pt40}
H.~G. Katzgraber and A.~P. Young, {\em Phys. Rev. B}, 2002, {\bf 65}, 214402.

\bibitem{pt39}
H.~G. Katzgraber and A.~P. Young, {\em Phys. Rev. B}, 2002, {\bf 65}, 214401.

\bibitem{pt116}
H.~G. Katzgraber, {\em Comp. Phys. Comm.}, 2002, {\bf 147}, 439.

\bibitem{pt121}
Y.~Ishikawa, Y.~Sujita, T.~Nishikawa, and Y.~Okamoto, {\em Chem. Phys. Lett.},
  2001, {\bf 333}, 199.

\bibitem{pt38}
P.~Sengupta, A.~W. Sandvik, and D.~K. Campbell, {\em Phys. Rev. B}, 2002, {\bf
  65}, 155113.

\bibitem{pt2}
S.~Shin, W.~Son, and S.~Jang, {\em Theochem}, 2004, {\bf 673}, 109.

\bibitem{pt140}
F.~Calco, F.~Spiegelman, M.~A. Gaveau, M.~Briant, P.~R. Fournier, J.~M.
  Mestdagh, and J.~P. Visticot, {\em Euro. Phys. J. D}, 2003, {\bf 24}, 215.

\bibitem{pt168}
G.~E. Lopez, {\em J. Chem. Phys.}, 2002, {\bf 117}, 2225.

\bibitem{pt52}
M.~Habeck, M.~Nilges, and W.~Rieping, {\em Phys. Rev. Lett.}, 2005, {\bf 94},
  018105.

\bibitem{pt141}
C.~Bertrand, M.~Ohmi, R.~Suzuki, and H.~Kado, {\em IEEE Trans. Biomed. Eng.},
  2001, {\bf 48}, 533.

\bibitem{pt123}
H.~Merlitz and W.~Wenzel, {\em Chem. Phys. Lett.}, 2002, {\bf 362}, 271.

\bibitem{pt136}
H.~Kozumi, {\em Comp. Stat. Data Anal.}, 2004, {\bf 46}, 441.

\bibitem{EarlDeem}
D.~J. Earl and M.~W. Deem, {\em J. Phys. Chem. B}, 2004, {\bf 108}, 6844.

\bibitem{XPLOR}
A.~T. Brunger, J.~Kuriyan, and M.~Karplus, {\em Science}, 1987, {\bf 235}, 458.

\bibitem{Shakhnovich}
R.~S. DeWitte and E.~I. Shakhnovich, {\em J. Am. Chem. Soc.}, 1996, {\bf 118},
  11733.

\bibitem{Fredrickson}
A.~Alexander-Katz, A.~G. Moreira, S.~W. Sides, and G.~H. Fredrickson, {\em J.
  Chem. Phys.}, 2005, {\bf 122}, 014904.

\bibitem{pt203}
M.~Falcioni and M.~W. Deem, {\em Phys. Rev. E}, 2000, {\bf 61}, 5948.

\bibitem{BogaradDeem}
L.~D. Bogarad and M.~W. Deem, {\em Proc. Natl. Acad. Sci. USA}, 1999, {\bf 96},
  2591.

\end{thebibliography}

\clearpage

\begin{figure}[ht]
  \begin{center}
   \epsfig{file=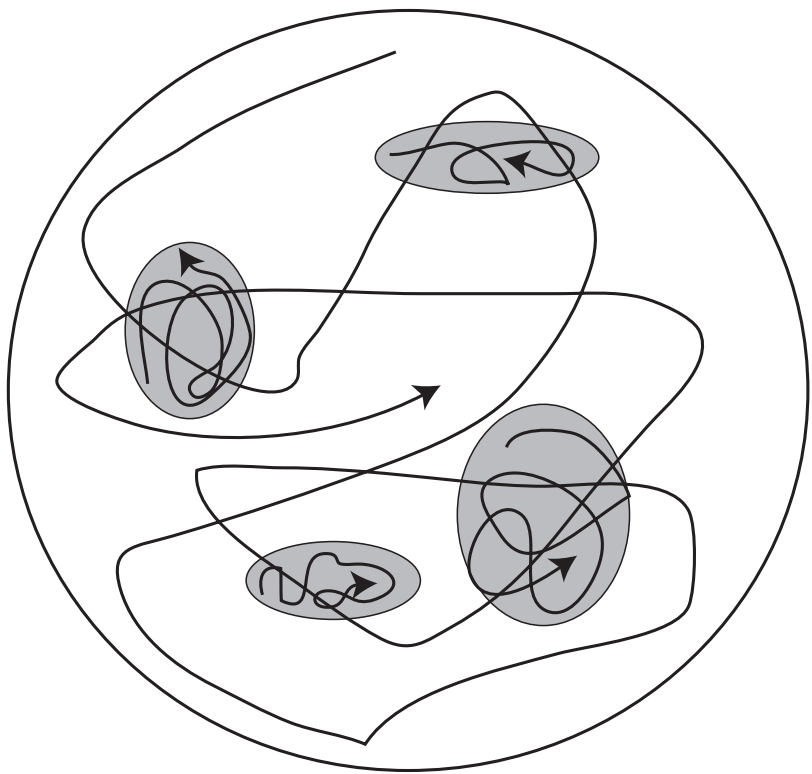,height=4in,clip=,angle=0}
   \caption{2-D representation of phase space. A simulation at
lower temperatures can become trapped in a non-representative
sample of the low free energy minima
(shaded regions). At higher temperatures, a simulation can sample
more of phase space (light plus shaded areas). Configuration swaps
between the lower and higher temperature systems allow the lower
temperature systems to escape from one region of phase space where
they were effectively `stuck' and to sample a representative
set of the low free energy minima.}
  \end{center}
\end{figure}

\newpage
\begin{figure}[ht]
  \begin{center}
   \epsfig{file=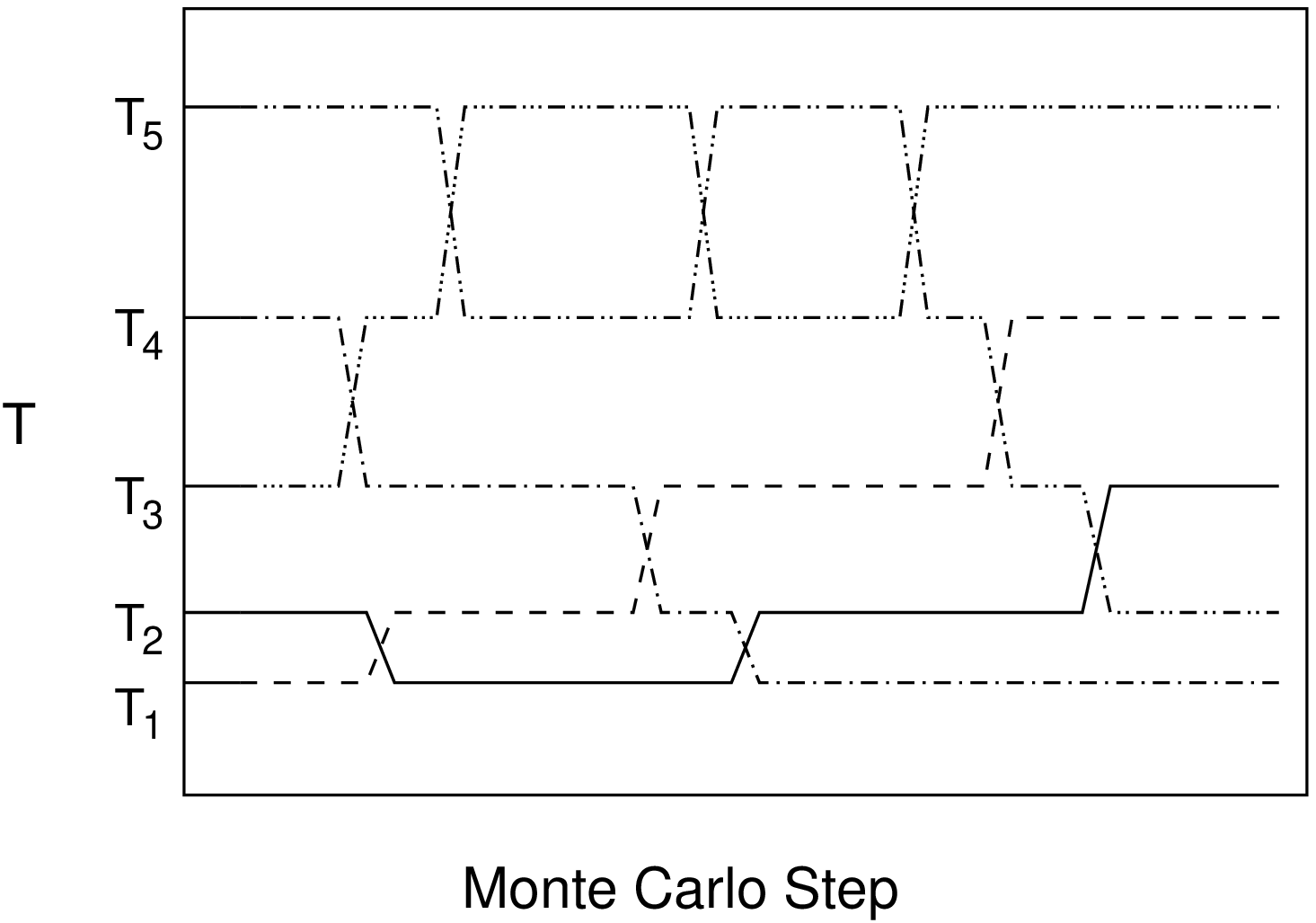,height=4in,clip=,angle=0}
   \caption{Schematic representation of parallel tempering swaps
between adjacent replicas at different temperatures.  In between
the swaps, several constant-temperature Monte Carlo moves are performed.}
  \end{center}
\end{figure}

\newpage
\begin{figure}[ht]
  \begin{center}
   \epsfig{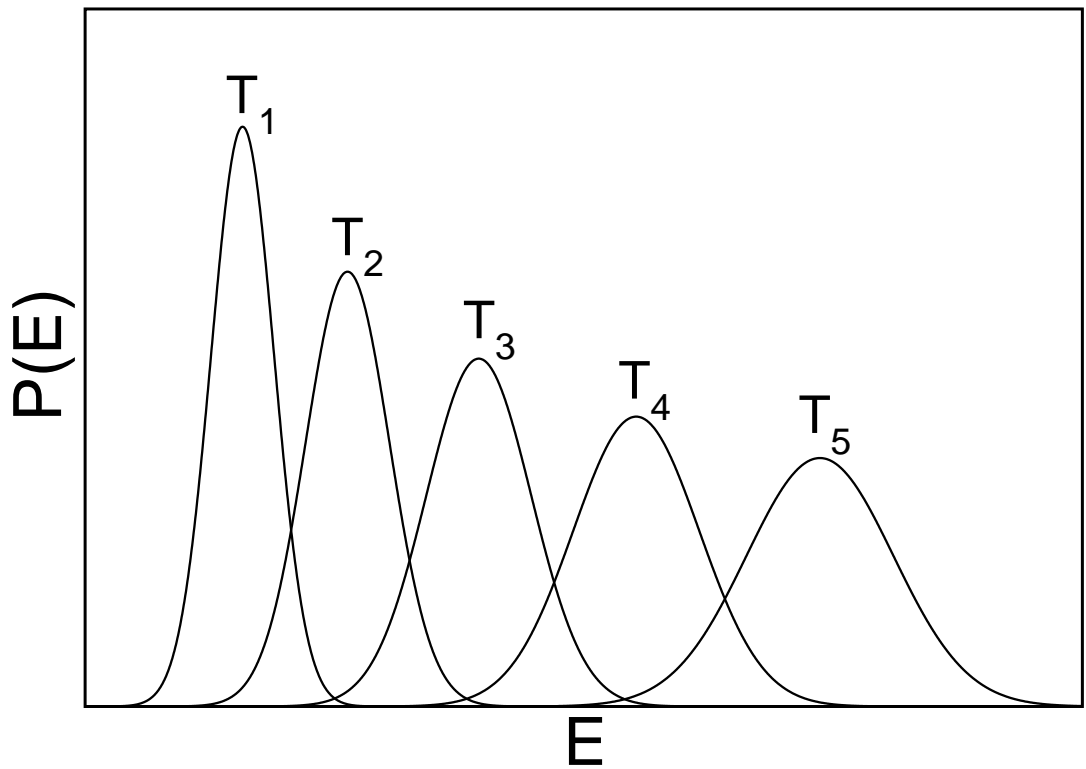}
   \caption{Energy histograms for a model system at five different
temperatures. Overlap of the energy histograms between adjacent 
replicas  at different temperatures allows for acceptance
of the configuration swaps.}
  \end{center}
\end{figure}

\newpage
\begin{figure}[ht]
  \begin{center}
   \epsfig{file=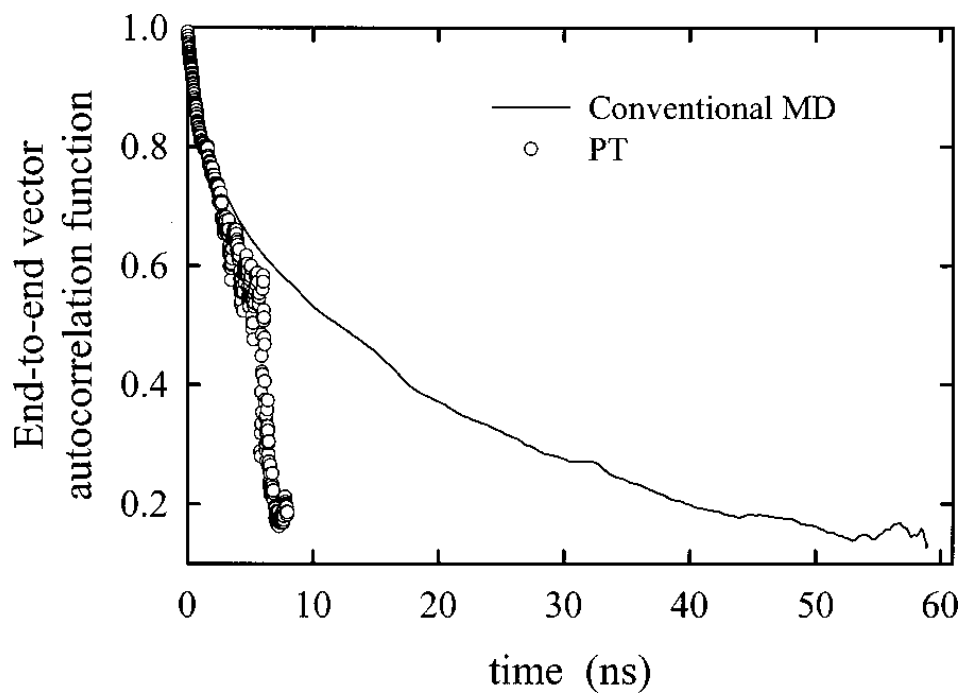,height=4in,clip=,angle=0}
   \caption{End-to-end correlation function for polymeric
1,4-polybutadiene.
The parallel tempering simulation relaxes much more quickly, and
is, thus, a more efficient simulation.
Used with permission.\cite{pt158}
Copyright 2001, American Institute of Physics.
}
  \end{center}
\end{figure}

\end{document}